\documentclass[prl,aps,twocolumn,showpacs]{revtex4}
\usepackage{graphicx,latexsym}
\usepackage{dcolumn}
\usepackage{amsmath,amssymb,epsf,bm}
\begin{document}
\title{Spin-Hall effect on edge magnetization
and electric conductance of a 2D semiconductor strip}
\author{A.~G. Mal'shukov$^1$, L.~Y. Wang$^{2}$, C.~S. Chu$^{2}$ and K.~A. Chao$^3$}
\affiliation{$^1$Institute of Spectroscopy, Russian Academy of
Science, 142190, Troitsk, Moscow oblast, Russia \\
$^2$Department of Electrophysics, National Chiao Tung
University, Hsinchu 30010, Taiwan \\
$^3$ Solid State Theory Division, Department of Physics, Lund
University, S-22362 Lund, Sweden}
\begin{abstract}
The intrinsic spin-Hall effect on spin accumulation and electric
conductance in a diffusive regime of a 2D electron gas has been
studied for a 2D strip of a finite width. It is shown that the
spin polarization near the flanks of the strip, as well as the
electric current in the longitudinal direction exhibit damped
oscillations as a function of the width and strength of the
Dresselhaus spin-orbit interaction. Cubic terms of this
interaction are crucial for spin accumulation near the edges. As
expected, no effect on the spin accumulation and electric
conductance have been found in case of Rashba  spin-orbit
interaction.
\end{abstract}

\maketitle

Spintronics is a fast developing area to use electron spin degrees
of freedom in electronic devices~\cite{Prinz}. One of its most
challenging goals is to find a method for manipulating electron
spins by electric fields. The spin-orbit interaction (SOI), which
couples the electron momentum and spin, can be a mediator between
the charge and spin degrees of freedom. Such a coupling gives rise
to the so called spin-Hall effect (SHE) which attracted much
interest recently. Due to SOI the spin flow can be induced
perpendicular to the DC electric field, as has been predicted for
systems containing spin-orbit impurity scatterers \cite{Hirsch}.
Later, similar phenomenon was predicted for noncentrosymmetric
semiconductors with spin split electron and hole energy bands
~\cite{Murakami}. It was called the {\it intrinsic} spin-Hall
effect, in contrast to the {\it extrinsic} impurity induced
effect, because in the former case it originates from the
electronic band structure of a semiconductor sample. Since the
spin current carries the spin polarization, one would expect a
buildup of the spin density near the sample boundaries. In fact,
this accumulated polarization is a first signature of SHE which
has been detected experimentally, confirming thus the extrinsic
SHE ~\cite{Awschalom2} in semiconductor films and intrinsic SHE in
a 2D hole gas ~\cite{Wunderlich}. On the other hand, there were
still no experimental evidence of intrinsic SHE in 2D electron
gases. The possibility of such an effect in macroscopic samples
with a finite elastic mean free path of electrons caused recently
much debates. It has been shown analytically
~\cite{Inoue+,Mischenko,Dimitrova,Burkov,Raimondi,MalshDress} and
numerically ~\cite{Nomura} that in such systems SHE vanishes at
arbitrary weak disorder in DC limit, for isotropic, as well as
anisotropic \cite{Raimondi} impurity scattering, when  SOI is
represented by the so called Rashba interaction ~\cite{rash}. As
one can expect in this case, there is no spin accumulation at the
sample boundaries, except for the pockets near the electric
contacts ~\cite{Mischenko}. At the same time, the Dresselhaus SOI
~\cite{dress}, which dominates in symmetric quantum wells, gives a
finite spin-Hall conductivity ~\cite{MalshDress}. The latter can
be of the order of its universal value $e/8\pi \hbar$. The same
has been shown for the cubic Rashba interaction in hole systems
~\cite{Nomura,comment}. In this connection an important question
is what sort of the spin accumulation could Dresselhaus SOI induce
near sample boundaries. Another problem which, as far as we know,
was not discussed in literature, is how the {\it electric} current
along the applied electric field will change under SHE. In the
present work we will use the diffusion approximation for the
electron transport to derive the drift diffusion equations with
corresponding boundary conditions for the spin and charge
densities coupled to each other via SOI of general form. Then the
spin density near the flanks of an infinite 2D strip and the
correction to its longitudinal electric resistance will be
calculated for Dresselhaus and Rashba SOI.

Let us consider 2DEG confined in an infinite 2D strip. The
boundaries of the strip are at $y=\pm d/2$. The electric field $E$
drives the DC current in the $x$-direction and induces the
spin-Hall current in the $y$-direction. This current leads to spin
polarization buildup near boundaries. Since $d \gg k_F^{-1}$,
where $k_F$ is the Fermi wavector, this problem can be treated
within the semiclassical approximation. Moreover, we will assume
that $d$ is much larger than the electron elastic mean free path
$l$, so that the drift-diffusion equation can be applied for
description of the spin and charge transport. Our goal is to
derive this equation for SOI of general form
\begin{equation}\label{Hso}
H_{so} = \bm{h}_{\bm{k}}\cdot\bm{\sigma} \, ,
\end{equation}
where $\bm{\sigma}$$\equiv$$(\sigma^x,\sigma^y,\sigma^z)$ is the
Pauli matrix vector, and the effective magnetic field
$\bm{h}_{\bm{k}}=-\bm{h}_{-\bm{k}}$ is a function of the
two-dimensional wave-vector $\bm{k}$.

We start from determining linear responses to the magnetic
$\bm{B}(\bm{r},t)$ and electric $V(\bm{r},t)$ potentials. The
magnetic potentials are introduced in order to derive the
diffusion equation and play an auxiliary role. The corresponding
one-particle interaction with the spin density is defined as $
H_{sp} = \bm{B}(\bm{r},t)\cdot\bm{\sigma}$. These potentials
induce the spin and charge densities, $\bm{S}(\bm{r},t)$ and
$n(\bm{r},t)$, respectively. Due to SOI the charge and spin
degrees of freedom are coupled, so that the electric potential can
induce the spin density \cite{Edelstein} and \emph{vice versa}.
Therefore, it is convenient to introduce the four-vector of
densities $D_i(\bm{r},t)$, such that $D_0(\bm{r})=n(\bm{r},t)$ and
$D_{x,y,z}(\bm{r},t)=S_{x,y,z}(\bm{r},t)$. The corresponding
four-vector of potentials will be denoted as $\Phi_i(\bm{r},t)$.
Accordingly, the linear response equations can be written in the
form
\begin{eqnarray}\label{response}
D_i(\bm{r},t)&=&\int d^2r^{\prime} dt^{\prime}\sum_j
\Pi_{ij}(\bm{r},\bm{r}^{\prime},t-t^{\prime})\Phi_j(\bm{r}^{\prime},t^{\prime})\nonumber\\
&+& D_i^0(\bm{r},t)\,.
\end{eqnarray}
The response functions
$\Pi_{ij}(\bm{r},\bm{r}^{\prime},t-t^{\prime})$ can be expressed
in a standard way \cite{agd} through the retarded and advanced
Green functions $G^r(\bm{r},\bm{r}^{\prime},t)$ and
$G^a(\bm{r},\bm{r}^{\prime},t)$. In the Fourier representation we
get
\begin{eqnarray}\label{pi}
\Pi_{ij}(\bm{r},\bm{r}^{\prime},\omega)&=&i\omega \int
\frac{d\omega^{\prime}}{2\pi} \frac{\partial
n_F(\omega^{\prime})}{\partial\omega^{\prime}}\langle
Tr[G^a(\bm{r}^{\prime},\bm{r},\omega^{\prime}) \times \nonumber \\
&& \times \Sigma_i
G^r(\bm{r},\bm{r}^{\prime},\omega^{\prime}+\omega)\Sigma_j]
\rangle \,,
\end{eqnarray}
where  $\Sigma_0=1$, $\Sigma_i=\sigma_i$ at $i=x,y,z$ and
$n_F(\omega)$ is the Fermi distribution function. The time Fourier
components of densities $D_i^0(\bm{r},t)$ at $\omega \ll E_F$ are
defined as
\begin{eqnarray}\label{D0}
D_i^0(\bm{r},\omega)&=&i \int d^2r^{\prime}\sum_j
\Phi_j(\bm{r}^{\prime},\omega)\int
\frac{d\omega^{\prime}}{2\pi} n_F(\omega^{\prime})\nonumber \\
&\times& \langle Tr[G^r(\bm{r},\bm{r}^{\prime},\omega^{\prime})
\Sigma_i G^r(\bm{r}^{\prime},\bm{r},\omega^{\prime})\Sigma_j \nonumber \\
&-&G^a(\bm{r},\bm{r}^{\prime},\omega^{\prime}) \Sigma_i
G^a(\bm{r}^{\prime},\bm{k},\omega^{\prime})\Sigma_j] \rangle \,.
\end{eqnarray}
The trace in Eqs. (\ref{pi}-\ref{D0}) runs through the spin
variables, and the angular brackets denote the average over the
random distribution of impurities. Within the semiclassical
approximation the average of the product of Green functions can be
calculated perturbatively. Ignoring the weak localization effects,
the perturbation expansion consists of the so called ladder
series~\cite{agd,alt}. At small $\omega$ and large
$|\bm{r}-\bm{r}^{\prime}|$ they describe the particle and spin
diffusion processes. The building blocks for the perturbation
expansion are the average Green functions $\mathcal{G}^r$ and
$\mathcal{G}^a$, together with the pair correlator of the impurity
scattering potential $U_{sc}(\bm{r})$. A simple model of the
short-range isotropic potential gives $\langle
U_{sc}(\bm{r})U_{sc}(\bm{r}^{\prime})\rangle=\Gamma
\delta(\bm{r}-\bm{r}^{\prime})/\pi N_0$, where $N_0$ is the
electron density of states at the Fermi energy and
$\Gamma=1/2\tau$. Within the semiclassical approach the explicit
behavior of the electron wave functions near the boundaries of the
strip is not important. Therefore, the bulk expressions can be
used for the average Green functions. Hence, in the plane wave
representation
\begin{equation}\label{Green}
\mathcal{G}^r(\bm{k},\omega) =
[\mathcal{G}^{a}(\bm{k},\omega)]^{\dag} = (\omega - E_{\bm{k}} -
\bm{h}_{\bm{k}}\cdot\bm{\sigma} + i\Gamma)^{-1} \, ,
\end{equation}
where $E_{\bm{k}}$=$k^2/(2m^*)-E_F$. Since the integral in
(\ref{D0}) rapidly converges at $|\bm{r}-\bm{r}^{\prime}|\lesssim
k_F^{-1}$, $D_i^0(\bm{r},\omega)$ are given by the local values of
potentials. From (\ref{D0})-(\ref{Green}) one easily obtains the
local equilibrium densities
\begin{equation}\label{D0local}
D_i^0(\bm{r},\omega)=-2N_0 \Phi_i(\bm{r},\omega)\,.
\end{equation}
In their turn, the nonequlibrium spin and charge densities are
represented by the first term in Eq.~(\ref{response}). Within the
diffusion approximation this term is given by the gradient
expansion of (\ref{pi}) \cite{alt}. Such an expansion is valid as
far as spatial variations of $D_i(\bm{r},\omega)$ are relatively
small within the length of the order of the mean free path $l$.
The length scale for spin density variations near the boundaries
of the strip is given by $v_F/h_{\bm{k}_F}$. Hence, the diffusion
approximation can be employed only in the dirty limit
$h_{\bm{k}_F} \ll 1/\tau$. The diffusion equation is obtained
after the ladder summation in the first term of  Eq.~(\ref{pi})
and multiplying this equation by the operator inverse to
$\Pi_{ij}(\bm{r},\bm{r}^{\prime},\omega)$, as it has been
previously done in \cite{MalshDiff,MalshWL}. After some algebraic
manipulations one gets
\begin{equation}\label{diffusion}
\sum_j \mathcal{D}^{ij}(D_j-D^0_j)=-i\omega D_i \,,
\end{equation}
where the diffusion operator $\mathcal{D}^{ij}$ can be written as
\begin{equation}\label{diffusion2}
\mathcal{D}^{ij}=\delta^{ij}
D\bm{\nabla}^2-\Gamma^{ij}+R^{ijm}\nabla_m+M^{ij}\,.
\end{equation}
The first term represents the usual diffusion of the spin and
charge densities, while the second one describes the
D'akonov-Perel' \cite{dp} spin relaxation
\begin{equation}\label{DP}
\Gamma^{ij} = 4\tau \, \overline{h^2_{\bm{k}}[\delta^{ij}  -
n^i_{\bm{k}} n^j_{\bm{k}}]} \,,
\end{equation}
where $i,j\neq 0$, the overline denotes the average over the Fermi
surface and ${\bm{n}}_{\bm{k}}={\bm{h}}_{\bm{k}}/h_{\bm{k}}$. The
third term gives rise to precession of the inhomogeneous spin
polarization in the effective field of SOI \cite{MalshDiff}
\begin{equation}\label{precession}
R^{ijm}=4\tau\sum_l\varepsilon^{ijl}\, \overline{h_{\bm{k}}^{l}
v_F^m}\,.
\end{equation}
The nondiagonal elements of the form $\mathcal{D}^{i0}$ appear due
to spin-orbit mixing of spin and charge degrees of freedom. They
are collected in $M^{ij}$. For Rashba SOI $M^{i0}$ have been
calculated in \cite{Mischenko,Burkov}. In general case
\begin{equation}\label{M}
M^{i0}=\overline{\frac{h^3_{\bm{k}}}{\Gamma^2}\frac{\partial
n^i_{\bm{k}}}{\partial \bm{k}}}\cdot \bm{\nabla} \, .
\end{equation}

When a time independent homogeneous electric field is applied to
the system one has $\Phi_0=eEx$ and $D_0^0=-2N_0eEx$. At the same
time, $\Phi_i=0$ and, hence, $D_i^0=0$ at $i=x,y,z$. Due to charge
neutrality the induced charge density $eD_0=0$. It should be noted
that in the system under consideration the charge neutrality can
not be fulfilled precisely. The spin polarization accumulated at
the strip boundaries gives rise to charge accumulation via the
$M^{0i}$ terms in (\ref{diffusion})-(\ref{diffusion2}). The
screening effect will, however, strongly reduce this additional
charge, because the screening length of 2DEG is much less than the
typical length scale of spin density variations. We will ignore
such a small correction and set $D_0=0$ in (\ref{diffusion}). In
this way one arrives to the closed diffusion equation for three
components of the spin density. This equation coincides with the
usual equation describing diffusive propagation of the spin
density \cite{MalshDiff}, for exception of the additional term
$-M^{i0}D_0^0=2N_0eE\overline{h^3_{\bm{k}}
\nabla^x_{\bm{k}}n^i_{\bm{k}}}/\Gamma^2$ due to the external
electric field. Its origin becomes more clear in an infinite
system where the spin density is constant in space and only
$\Gamma^{ij}$ and $M^{ij}$ are retained in
(\ref{diffusion})-(\ref{diffusion2}). Hence, the solution of
(\ref{diffusion2}) at $\omega=0$ is $S_i \equiv S_i^{\text{b}}$,
with
\begin{equation}\label{S}
S_i^{\text{b}} \equiv
D_i^0/2=\frac{N_0eE}{\Gamma^2}\sum_j(\Gamma^{-1})^{ij}\overline{h^3_{\bm{k}}
\frac{\partial n^j_{\bm{k}}}{\partial k_x}}\,,
\end{equation}
where $(\Gamma^{-1})^{ij}$ is the matrix inverse to (\ref{DP}).
Such a phenomenon of spin orientation by the electric field was
predicted in Ref.~\cite{Edelstein} and recently observed in
\cite{Kato2}. In the special case of Rashba SOI
$\bm{h}_{\bm{k}}=\alpha(k) (\bm{k}\times \bm{z})$ it is easily to
get from (\ref{S}) the result of Ref.~\cite{Edelstein}
$S_y^{\text{b}}= -N_0eE\alpha\tau$.

In addition to the diffusion equation one needs the boundary
conditions. These conditions are that the three components of the
spin flux $I^y_x, I^y_y, I^y_z$ flowing in the $y$ direction turn
to 0 at $y=\pm d/2$. The linear response theory, similar to
(\ref{response}) gives
\begin{equation}\label{crresponse}
I^l_i(\bm{r},t)=\int d^2r^{\prime} dt^{\prime}\sum_j
\Xi^l_{ij}(\bm{r},\bm{r}^{\prime},t-t^{\prime})\Phi_j(\bm{r}^{\prime},t^{\prime})\,,
\end{equation}
where the response function $\Xi$ is given by
\begin{eqnarray}\label{xi}
\Xi^l_{ij}(\bm{r},\bm{r}^{\prime},\omega)&=&i\omega \int
\frac{d\omega^{\prime}}{2\pi} \frac{\partial
n_F(\omega^{\prime})}{\partial\omega^{\prime}}\langle
Tr[G^a(\bm{r}^{\prime},\bm{r},\omega^{\prime}) \times \nonumber \\
&& \times J^l_i
G^r(\bm{r},\bm{r}^{\prime},\omega^{\prime}+\omega)\Sigma_j]
\rangle \,,
\end{eqnarray}
with the one-particle spin-current operator defined by
$J^l_i$=($\sigma^i\mathrm{v}_l$+$\mathrm{v}_l\sigma^i$)/4 and the
particle velocity
\begin{equation}\label{v}
\mathrm{v}_l = \frac{k_l}{m^*} + \frac{\partial}{\partial
k_l}(\bm{h}_{\bm{k}}\cdot\bm{\sigma})\, .
\end{equation}
Taking into account (\ref{diffusion}-\ref{D0local}), we obtain
from (\ref{crresponse}-\ref{xi})
\begin{equation}\label{I}
 I^y_i(\bm{r})=-D\frac{\partial S_i}{\partial y} -
 \frac{1}{2}R^{ijy}(S_j-S_j^{\text{b}})+\delta_{iz}I_{sH}\,.
\end{equation}
The first two terms represent the diffusion spin current and the
current associated with the spin precession. The third term is the
uniform spin-Hall current polarized along the z axis. It is given
by
\begin{equation}\label{IsH}
I_{sH}= -\frac{1}{2}R^{zjy}S_j^{\text{b}}+eE \frac{N_0}{\Gamma^2}
\overline{v_F^y (\frac{\partial \bm{h}_{\bm{k}}}{\partial k^x}
\times \bm{h}_{\bm{k}})_z}\,.
\end{equation}
From (\ref{precession},\ref{S}) it is easily to see that for
Rashba SOI both terms in (\ref{IsH}) cancel each other making
$I_{sH}=0$, in accordance with
\cite{Inoue+,Mischenko,Dimitrova,Burkov,Raimondi,MalshDress,Nomura}.
Therefore, in case of the strip the solution of the diffusion
equation satisfying the boundary condition is $S_j=\delta_{jy}
S_y^{\text{b}}$. Hence, the spin density is uniform and does not
accumulate near boundaries. It should be noted that such
accumulation can, however, take place in the ballistic regime of
electron scattering \cite{ballistics}. At the same time, as shown
in Ref.~\cite{MalshDress}, even in the diffusive regime $I_{sH}
\neq 0$ for the Dresselhaus SOI. This inevitably leads to the spin
accumulation. Taking Dresselhaus SOI in the form
\begin{equation}\label{hD}
h^x_{\bm{k}} = \beta k_x (k^2_{y} - \kappa^2) \,\,\,;~h^y_{\bm{k}}
= -\beta k_y (k^2_{x} - \kappa^2) \, ,
\end{equation}
one can see that the bulk spin polarization (\ref{S}) has a
nonzero $S_x^{\text{b}}$ component, $R^{zxy}\neq 0$, while
$R^{zyy}=0$. Hence, the solution of the diffusion equation
(\ref{diffusion}) with the boundary condition $I_x^y(\pm
d/2)=I_z^y(\pm d/2)=0$ is $S_x, S_z \neq 0$ $S_y=0$. Let us define
$\Delta S_i(y)=S_i(y)- S_i^{\text{b}}$. The dependence of $\Delta
S_i(\pm d/2)$ from the strip width, as well as an example of
$\Delta S_z$ coordinate dependence, are shown in Fig.~1.
\begin{figure}[tbp]
\includegraphics[width=7.5cm, height=7.0cm]{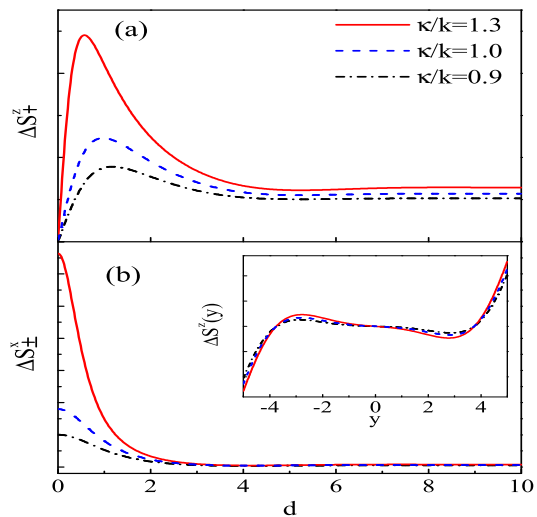}
\caption{Spin densities $\Delta S_i(\pm d/2)\equiv \Delta
S^i_{\pm}$ for $i=x,z$ on the boundaries of the strip, as
functions of its width $d$, for $\kappa/k$=0.9,1.0, and 1.3,
respectively. The insert shows the dependence of $\Delta S_z(y)$
on the transverse coordinate $y$. Lengths are measured in units of
$l_{so}=v_F^2 \hbar/2(\overline{v_{Fy}h_{\bm{k}y}})$} \label{fig1}
\end{figure}
The damped oscillation in the $d$-dependence of the spin
accumulation on the flanks of the strip can be seen for the $S_z$
polarization. Similar oscillations take place also in the
coordinate dependence. The length scale of these oscillations is
determined by the spin precession in the effective spin-orbit
field.

The arbitrary units have been used in Fig.~1. For a numerical
evaluation let us take $E=10^4$V/m,
$\sqrt{\overline{h^2_{\bm{k}_F}}} \tau/\hbar=0.1$, and
$\kappa/k_F$=0.8 for a GaAs quantum well of the width $w$=100\AA
 doped with $1.5\cdot 10^{15}$m$^{-2}$ electrons. We thus obtain
$|\Delta S_z(\pm d/2)|\simeq 5 \times 10^{11}$m$^{-2}$. The
corresponding volume density $\Delta S_z/w \simeq 5 \times
10^{19}$m$^{-3}$, which is within the sensitivity range of the
Faraday rotation method \cite{Awschalom2}.

It should be noted that in the considered here "dirty" limit
$\sqrt{\overline{h^2_{\bm{k}_F}}} \tau/\hbar \ll 1$ the spin-Hall
current is suppressed by the impurity scattering. As shown in
\cite{MalshDress,Nomura} for Dresselhaus and cubic Rashba SOI,
this current decreases as $\overline{h^2_{\bm{k}_F}}
\tau^2/\hbar^2$ down from its highest universal value. At the same
time, an analysis of the diffusion equation shows that the
accumulated at the flanks of the strip spin density decreases
slower, as $\sqrt{\overline{h^2_{\bm{k}_F}}} \tau/\hbar$. This
explains why for the considered above realistic numerical
parameters, even in the dirty case, the noticeable spin
polarization can be accumulated near the boundary.

Usually, the spin-Hall effect is associated with the spin
polarization flow, or the spin density accumulation on the sample
edges, in response to the electric field. On the other hand, this
effect can show up in the \emph{electric } conductance as well. To
see such an effect we take 0-projection of (\ref{crresponse}),
which by definition is the electric current. The current flows
along the $x$ axis. The corresponding response function
$\Xi^x_{0j}$ is given by (\ref{xi}) with $J^x_0=\mathrm{v}_x$.
Using Eqs. (\ref{xi}), (\ref{v}), (\ref{diffusion}) and expressing
$\Phi_i$ from (\ref{D0local}) one gets the electric current
density
\begin{equation}\label{Ielectric}
I^x=\sigma E + A \frac{\partial S_z}{\partial y} \,,
\end{equation}
where $\sigma$ is the Drude conductivity and
\begin{equation}\label{A}
A=e \frac{1}{2\Gamma^2} \left [2\overline{v_F^y (\frac{\partial
\bm{h}_{\bm{k}}}{\partial k^x} \times
\bm{h}_{\bm{k}})_z}+\overline{v_F^x (\frac{\partial
\bm{h}_{\bm{k}}}{\partial k^y} \times \bm{h}_{\bm{k}})_z}\right
]\,.
\end{equation}
The total current is obtained by integrating (\ref{Ielectric})
over $y$. Therefore, the spin-Hall correction to the strip
conductance
\begin{equation}\label{G}
\Delta
G=\frac{A}{E}\left(S_z(d/2)-S_z(-d/2)\right)=\frac{2A}{E}S_z(d/2)\,.
\end{equation}
Hence, the dependence of $\Delta G$ on the strip width coincides
with that of the spin density shown in Fig.~1a.

In conclusion, we employed the diffusion approximation to study
the spin-Hall effect in an infinite 2D strip. In case of the
Dresselhaus spin-orbit interaction this effect leads to spin
accumulation near the flanks of the strip, as well as to a
correction to the longitudinal electric conductance. Both, the
spin accumulation and the conductance exhibit damped oscillations
as a function of the strip width.

This work was supported by the Taiwan National Science Council
NSC93-2112-M-009-036 and  RFBR Grant No 030217452.


\begin{thebibliography}{99}
\bibitem{comment}
The cubic Rashba interaction should not be confused with the
conventional linear Rashba SOI with the wave-vector dependent
coupling constant $\alpha(k)$. In the latter case SHE
$\propto$$\alpha(k)k/E_F \ll 1$ \cite{Nomura,Dimitrova}.
\end{thebibliography}
\end{document}